\documentclass{article}

\usepackage{tikz}
\usetikzlibrary{shapes,calc,decorations.markings,decorations.pathreplacing,
arrows,chains,shapes.multipart,shapes,fit,automata,positioning,arrows.meta,
patterns, decorations.pathmorphing}

\usepackage{caption}
\usepackage{algorithm}
\usepackage{algpseudocode}
\usepackage{pifont}
\usepackage{epsfig}
\usepackage{tabularx}
\usepackage{multirow}
\usepackage{graphicx}
\usepackage{subcaption}
\usepackage{hyperref}
\usepackage{forest}

\usepackage{amssymb, amsmath}

\newcommand{\x}{\mathbf{x}}

\newcommand{\RR}{\ensuremath{\mathbb{R}}}

\newcommand{\N}{\ensuremath{N}}

\begin{document}

\title{Computing Tropical Prevarieties in Parallel\thanks{This material 
is based upon work
supported by the National Science Foundation under Grant No. 1440534.}}

\date{23 June 2017}

\author{Anders Jensen\thanks{Aarhus University} \and
Jeff Sommars\thanks{University of Illinois at Chicago} \and
Jan Verschelde\thanks{University of Illinois at Chicago}}

\maketitle

\begin{abstract}
The computation of the tropical prevariety is the first step
in the application of polyhedral methods to compute positive
dimensional solution sets of polynomial systems.
In particular, pretropisms are candidate leading exponents
for the power series developments of the solutions.
The computation of the power series may start as soon as one
pretropism is available, so our parallel computation of the
tropical prevariety has an application in a pipelined solver.

We present a parallel implementation of dynamic enumeration.
Our first distributed memory implementation with forked processes
achieved good speedups, but quite often resulted in large variations
in the execution times of the processes.
The shared memory multithreaded version applies work stealing
to reduce the variability of the run time.
Our implementation applies the thread safe Parma Polyhedral Library (PPL),
in exact arithmetic with the GNU Multiprecision Arithmetic Library (GMP),
aided by the fast memory allocations of TCMalloc.

Our parallel implementation is capable of computing the tropical
prevariety of the cyclic 16-roots problem.
We also report on computational experiments on the $n$-body
and $n$-vortex problems;
our computational results compare favorably with Gfan.
\end{abstract}

\maketitle

\section{Introduction}

Given one polynomial in two variables,
the Newton-Puiseux algorithm computes series expansions
for the algebraic curve defined by the polynomial,
departing from the edges of the Newton polyhedron.
Given a polynomial system, the rays in the tropical prevariety
may lead to the series expansions for the
positive dimensional solution sets of the system.
This paper presents a parallel algorithm 
to compute the tropical prevariety.
We refer to~\cite{MS15} for a textbook introduction
to tropical algebraic geometry.

The {\em Newton polytope} $\textup{NP}(f)$ of a polynomial $f$ 
in $n$ variables is the convex hull of the exponent vectors appearing 
with nonzero coefficient in $f$.
The exponent vectors have integer coordinates.
Each face of $\textup{NP}(f)$ has a normal cone. The set of all normal cones constitutes the \emph{normal fan} of $\textup{NP}(f)$ which is a polyhedral fan in $\RR^n$. The \emph{tropical hypersurface} $\textup{T}(f)$ is the subfan of non-maximal cones. Given a polynomial system as a tuple of polynomials, the \emph{tropical prevariety} that we wish to compute is the intersection of all tropical hypersurfaces of polynomials in the tuple.

The tropical prevariety is mainly a combinatorial object depending only on the Newton polytopes of the polynomials in the system, whereas the cancellation properties of the coefficients are captured by the \emph{tropical variety} of the polynomial ideal of the system, which we shall not consider here.

For a general introduction to polytopes, we refer to~\cite{Zie95}.
In~\cite{BJSST07}, the tropical prevariety was defined via the
common refinement of normal fans. To be formally correct, we use the same definition.
Given two fans $F_1$ and $F_2$, their {\em common refinement} 
$F_1 \wedge F_2$ is defined as
\begin{equation} \label{eqdefrefinement1}
   F_1 \wedge F_2 =
   \{C_1 \cap C_2 | (C_1,C_2) \in F_1 \times F_2\}.
\end{equation}
As the common refinement of two fans is again a fan,
the common refinement of three fans $F_1$, $F_2$, and $F_3$
may be computed as $(F_1 \wedge F_2) \wedge F_3$.

The \emph{support} of a polyhedral fan is the union of its cones. 
We clarify the definition of the tropical prevariety of a tuple of 
polynomials $(f_1,f_2,\dots,f_\N)$ by defining it as the support 
of $T(f_1) \wedge T(f_2) \wedge \cdots\wedge T(f_\N)$. 
The nonzero vectors in the prevariety are called \emph{pretropisms}. 
The pretropisms are exactly the vectors normal to positive dimensional 
faces of each polytope in a tuple of Newton polytopes.

\noindent {\bf Problem Statement.}
Given a tuple of Newton polytopes \\ $(P_1, P_2, \ldots, P_\N)$
with normal fans $(F_1, F_2, \ldots, F_\N)$,
efficiently compute the tropical prevariety of the fans with a 
parallel implementation in order to compute more challenging examples.
Though the postprocessing of the cones of the prevariety is
embarassingly parallel, computing the tropical prevariety is not,
mainly because the cones in the output share parts of other cones.

\noindent {\bf Relation to Mixed Volumes.}
Our problem can be considered as a generalization of
the mixed volume computation.  For the relation between
triangulations and mixed subdivisions, we refer to~\cite{DRS10}. 
In the mixed volume computation, one intersects normal cones to
the edges of each polytope.  Linear programming models to prune
superfluous edge-edge combinations were proposed first in~\cite{EC95}.
Further developments can be found in~\cite{GL03} and~\cite{MTK07}.
A recent complexity study appeared in~\cite{Mal15},
along with a report on a parallel implementation of
the mixed volume computation.

Most relevant for the algorithms presented in this paper
is the dynamic enumeration introduced in~\cite{MTK07}.

\noindent {\bf Software.}
A practical study on various software packages for
exact volume computation of a polytope is described in~\cite{BEF00}.
The authors of~\cite{EF14} present an
experimental study of approximate polytope volume computation.
In~\cite{EFG16}, a total polynomial-time algorithm is presented to
compute the edge skeleton of a polytope.

Free dedicated software packages to compute mixed volumes
are MixedVol~\cite{GLW05} and DEMiCS~\cite{MT08}.
In~\cite{MT08}, the computation of the mixed volume for
the cyclic 16-roots problem was reported for the first time.
The mixed volume computation is included in PHCpack~\cite{Ver99},
pss5~\cite{Mal15}, and gfanlib~\cite{Jen16}.
Gfan~\cite{Jen08} contains software to compute the common refinement of
the normal fans of the Newton polytopes.
Gfan relies on cddlib~\cite{FP96} and optionally SoPlex~\cite{Wunderling1996} for lower level polyhedral computations.

The external software we used in our computations
is the Parma Polyhedral Library (PPL)~\cite{BHZ08},
and in particular its thread safe multithreading capabilities.
Speedups were obtained with TCMalloc~\cite{TCMalloc}.

\noindent {\bf Our contributions.}
This paper extends the results of the last two authors,
presented in~\cite{SV162,SV161},
extending the pruning algorithms with dynamic enumeration
and a work stealing strategy.
Our parallel implementation gives the first computation
of the tropical prevariety for the cyclic 16-roots problem.

\noindent {\bf Acknowledgments.}
We thank Enea Zaffanella for developing a 
thread safe version of PPL.

\section{Half Open Cones}

We represent the support of a fan as a set of mutually disjoint cones.
Such representations were also used in~\cite{CJR11}.
The key element in the representation
is the notion of a half open cone, which we will define next.

While a (closed) polyhedral cone in $\RR^n$ is a set of the form
\begin{equation} \label{eq:closedcones}
  \{ \ x \in\RR^n \ | \ Ax\leq 0 \ \}
\end{equation}
with $A\in\RR^{m\times n}$ and the comparison done coordinatewise,
a half open polyhedral cone is a set of the form
\begin{equation} \label{eq:halfopencones}
   \{ \ x\in\RR^n \ | \ Ax\leq 0 \wedge A'x< 0 \ \}
\end{equation}
with $A\in\RR^{m\times n}$ and $A'\in\RR^{m'\times n}$.

\subsection{Constructing Half Open Cones}

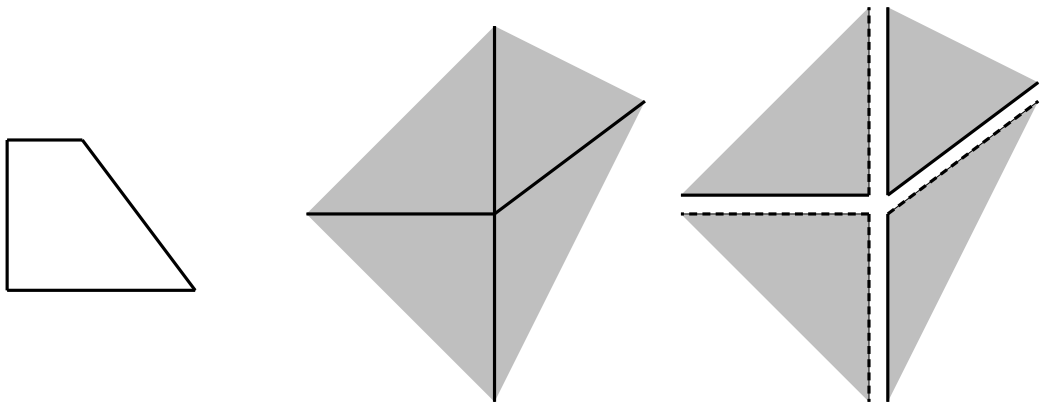
\begin{figure*}[htb]
\begin{minipage}[b]{.32\textwidth}
\centering
\begin{tikzpicture}[scale=.5]
\fill[draw=white] (-3,-3);
\fill[draw=black, very thick] (0,0) -- (-3,4);
\fill[draw=black, very thick] (-3,4) -- (-5,4);
\fill[draw=black, very thick] (-5,0) -- (-5,4);
\fill[draw=black, very thick] (0,0) -- (-5,0);
\end{tikzpicture}
\end{minipage}
\begin{minipage}[b]{.3\textwidth}
\begin{tikzpicture}[scale=.5]
\fill[fill = gray!50] (0,0) -- (-5,0) -- (0,-5) -- cycle;
\fill[fill = gray!50] (0,0) -- (-5,0) -- (0,5) -- cycle;
\fill[fill = gray!50] (0,0) -- (0,5) -- (4,3) -- cycle;
\fill[fill = gray!50] (0,0) -- (4,3) -- (0,-5) -- cycle;
\fill[draw=black, very thick] (0,0) -- (0,-5);
\fill[draw=black, very thick] (0,0) -- (4,3);
\fill[draw=black, very thick] (0,0) -- (0,5);
\fill[draw=black, very thick] (0,0) -- (-5,0);
\end{tikzpicture}
\end{minipage}
\begin{minipage}[b]{.3\textwidth}
\begin{tikzpicture}[scale=.5]
\def\yshift{0.5}
\def\xshift{0.5}

\fill[fill = gray!50] (0,0) -- (-5,0) -- (0,-5) -- cycle;
\fill[draw=black, very thick, densely dashed] (0,0) -- (0,-5);
\fill[draw=black, very thick, densely dashed] (0,0) -- (-5,0);

\fill[fill = gray!50] (0,\yshift) -- (-5,\yshift) -- (0,5 + \yshift) -- cycle;
\fill[draw=black, very thick, densely dashed] (0,\yshift) -- (0,5 + \yshift);
\fill[draw=black, very thick] (0,\yshift) -- (-5,\yshift);

\fill[fill = gray!50]
(\xshift,\yshift) -- (\xshift + 4,3+\yshift) -- (\xshift,5 + \yshift) -- cycle;
\fill[draw=black, very thick] (\xshift,\yshift) -- (\xshift,5 + \yshift);
\fill[draw=black, very thick] (\xshift,\yshift) -- (\xshift + 4,3+\yshift);

\fill[fill = gray!50] (\xshift,0) -- (\xshift + 4,3) -- (\xshift,-5) -- cycle;
\fill[draw=black, very thick] (\xshift,0) -- (\xshift,-5);
\fill[draw=black, very thick, densely dashed] (\xshift,0) -- (\xshift + 4,3);

\end{tikzpicture}
\end{minipage}

\caption{
{\it Left}: A Newton polytope $P$.
{\it Center}: The normal fan $F$ of $P$.
{\it Right}: $F$ split apart into four half open cones. The dashed
lines represent boundaries that are not contained by a cone, while the
solid lines represent
boundaries that are contained by a cone. In this example, the upper right cone
contains the origin.
}
\label{halfopencones}
\end{figure*}

We provide algorithms that divide the support of fans defined by a
convex polytope $P$ into a disjoint set of half open cones. We first explain how this is done for the normal fan of $P$ and later how it is done for the tropical hypersurface of $T(P)$ of $P$, by which we mean the tropical hypersurface of a polynomial with Newton polytope $P$.
Figure~\ref{halfopencones} shows an example of dividing the support of the normal fan of a single polytope into half open cones.

\begin{figure}[h]
\centering
\begin{tikzpicture}
	[x={(0.061125cm, -0.945586cm)},
	y={(-0.876347cm, 0.102421cm)},
	z={(0.477786cm, 0.308831cm)},
	scale=3.000000,
	back/.style={loosely dotted, thin},
	vertex/.style={inner sep=1pt,circle,draw=black!25!black,thick,anchor=base}
]

\node[vertex] (a) at (1.00000, 1.00000, 1.00000)     {7};
\node[vertex] (b) at (0.00000, 0.00000, 0.00000)     {2};
\node[vertex] (c) at (0.00000, 0.00000, 1.00000)     {1};
\node[vertex] (d) at (0.00000, 1.00000, 0.00000)     {6};
\node[vertex] (e) at (0.00000, 1.00000, 1.00000)     {5};
\node[vertex] (f) at (1.00000, 0.00000, 0.00000)     {4};
\node[vertex] (g) at (1.00000, 0.00000, 1.00000)     {3};
\node[vertex] (h) at (1.00000, 1.00000, 0.00000)     {8};
\draw[back,postaction={decorate, decoration={markings, mark=at position 0.5 with {\arrow{Latex[width=2mm]}}}}] (g) -- (a);
\draw[back,postaction={decorate, decoration={markings, mark=at position 0.5 with {\arrow{Latex[width=2mm]}}}}] (a) -- (h);
\draw[back,postaction={decorate, decoration={markings, mark=at position 0.7 with {\arrow{Latex[width=2mm]}}}}] (e) -- (a);
\draw[postaction={decorate, decoration={markings, mark=at position 0.5 with {\arrow{Latex[width=2mm]}}}}] (g) -- (f);
\draw[postaction={decorate, decoration={markings, mark=at position 0.5 with {\arrow{Latex[width=2mm]}}}}] (c) -- (b);
\draw[postaction={decorate, decoration={markings, mark=at position 0.5 with {\arrow{Latex[width=2mm]}}}}] (c) -- (e);
\draw[postaction={decorate, decoration={markings, mark=at position 0.5 with {\arrow{Latex[width=2mm]}}}}] (c) -- (g);
\draw[postaction={decorate, decoration={markings, mark=at position 0.5 with {\arrow{Latex[width=2mm]}}}}] (e) -- (d);
\draw[postaction={decorate, decoration={markings, mark=at position 0.5 with {\arrow{Latex[width=2mm]}}}}] (d) -- (h);
\draw[postaction={decorate, decoration={markings, mark=at position 0.5 with {\arrow{Latex[width=2mm]}}}}] (f) -- (h);
\draw[postaction={decorate, decoration={markings, mark=at position 0.7 with {\arrow{Latex[width=2mm]}}}}] (b) -- (d);
\draw[postaction={decorate, decoration={markings, mark=at position 0.5 with {\arrow{Latex[width=2mm]}}}}] (b) -- (f);

\draw[very thick, postaction={decorate, decoration={markings, mark=at position 1 with {\arrow{Latex[width=2mm]}}}}] (-0.5000,-0.5000,1.500) node[above] {$r$} -- (-.10000, -.10000, 1.10000);
\end{tikzpicture}
\caption{A cube oriented by a vector $r$.}
\label{cube}
\end{figure}
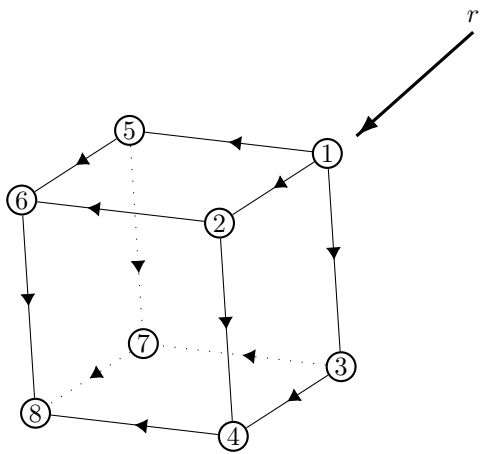

We begin by orienting
the edge graph of the polytope using a random vector $r\in\RR^n$ and then ordering the vertices by inner product with $r$. Assuming that these inner products are different, we find a unique sink orientation of the graph by giving
each edge of the polytope a direction based on the
vertex ordering, as in Figure~\ref{cube}.

Using the unique sink orientation, we define a half open normal cone to each vertex $v$ of $P$ in this way: for each edge $e$ incident to $v$, create an inequality from it,
making it strict if $e$ is outgoing and non-strict if it is ingoing. The collection of all cones obtained as $v$ varies divides $\RR^n$ into a disjoint union. For the cones to cover $\RR^n$, it is essential that the oriented graph is cycle free. This is guranteed by the construction.

To write the tropical hypersurface $T(P)$ as a disjoint union of half 
open cones, call Algorithm~\ref{alg:halfopencones} below on
each half open cone $C$ constructed above. 
The output is a collection of half open cones covering exactly 
the non-interior points of $C$. Taking the union of all these 
collections as $C$ varies results in a set of cones covering the support of $T(P)$ 
exactly and containing one cone for each edge of $P$.

\begin{algorithm}[hbt]
\begin{algorithmic}[5]
\caption{Create half open cones of codimension one from a full dimensional
half open cone.}
\label{alg:halfopencones}
\Require An inequality description of a full dimensional half open cone $C$
\Ensure A collection of disjoint half open cones with union equal to the boundary of $C$
\Function{CreateHalfOpenCones}{$C$}
\If{$C$ has only strict constraints}
\Return $\emptyset$
\Else
\State Choose a non-strict constraint $c$ of $C$
\State $C_<:= C \textup{ but with } c \textup{ being strict}$
\State $C_=:= C \textup{ but with } c \textup{ being an equation}$
\State \Return $C_=\cup \textup{CreateHalfOpenCones}(C_<)$
\EndIf
\EndFunction
\end{algorithmic}
\end{algorithm}

\subsection{Represention as a Closed Cone}

Linear programming involves sets of equations and non-strict inequalities 
defining closed polyhedra. Therefore we represent a half open cone $C$ 
defined by matrices $A$ and $A'$ from (\ref{eq:halfopencones}) by the 
closed cone $C'\subseteq\RR^{n+1}$ defined by the matrix 
\begin{equation}
   \begin{bmatrix}A&0\\A'&1\end{bmatrix}
\end{equation}
with a column of zeros and ones appended. Then $C=\pi(C'\cap(\RR^n\times\RR_{>0}))$ where $\pi:\RR^n\times\RR\rightarrow\RR^n$ denotes the projection forgetting the last coordinate. Observing that for all $x\in C$ we have $x\times\varepsilon\in C'$ for $\varepsilon>0$ sufficiently small, we obtain $\overline{C}\subseteq\pi(C'\cap(\RR^n\times\{0\}))$,
while the converse inclusion holds if and only if $C$ is non-empty. 
This happens if 
and only if $C'\cap(\RR^n\times\RR_{>0})\not=\emptyset$ and in that 
case $\textup{dim}(C')=\textup{dim}(C)+1$. 
Non-emptyness and other properties can be determined with linear programming.

\section{Static Enumeration}

The common refinement~(\ref{eqdefrefinement1}) in the definition of
the tropical prevariety has a constructive formulation:
compute the intersection of every combination of $\N$ cones, 
one from each fan, and add the non-empty intersections to the output.
This combinatorial algorithm requires $\prod_{i=1}^\N |P_i|$ cone 
intersections, where $|P_i|$ is the number of edges of polytope $P_i$.

The recursive formulation in Algorithm~\ref{static:recursive}
(defined as {\em static enumeration}) performs substantially 
fewer cone intersections.

\begin{algorithm}[H]
\begin{algorithmic}[5]
\Require A list $F$ of fans $F_1,\dots F_N$ in $\RR^n$ where each $F_i$ is represented by a list of cones covering the support of $F_i$.
\Ensure A list of cones covering the support of $F_1\wedge\cdots\wedge F_N$.
\Procedure{StaticEnumeration}{Cone $C$, Index $i$}
\If{$C\not=\emptyset$}
\If{$i>|F|$}
\State Output $C$
\Else
\For{each cone $D$ in $F_i$}
\State {\sc StaticEnumeration}($C\cap D$, $i+1$)
\EndFor
\EndIf
\EndIf
\EndProcedure
\State {\sc StaticEnumeration}($\RR^n$, 1)
\end{algorithmic}
\caption{Static enumeration}
\label{static:recursive}
\end{algorithm}

The recursive execution of Algorithm~\ref{static:recursive}
leads to a tree of cone intersections
with a default depth first traversal.
At the $i$th level of the tree are a set of cones
comprising the common refinement of the first $i$ fans.
If the cones in Algorithm~\ref{static:recursive} are closed, then the intersection $C$ is always non-empty. In particular one may want to remove duplicate cones at each level of the tree.
If the fans $F_i$ in Algorithm~\ref{static:recursive} each consist
of mutually disjoint half open cones, then the occurrence of
duplicate intersections is avoided and cone intersections can indeed be empty.
Thus, working with half open cones in Algorithm~\ref{static:recursive}
brings another substantial improvement.

\section{Dynamic Enumeration}

Inspired by~\cite{GL03} and~\cite{MTK07},
we reduce the number of cone intersections by dynamic enumeration.
Dynamic enumeration can be viewed as a greedy method to reorder 
the fans during the computation, thereby affecting the shape of the tree of cone intersections. 
A random permutation of the $\N$ polytopes before
the start of Algorithm~\ref{static:recursive}
defines a sub-optimal order in which to intersect cones.
Using a greedy metric, defined in Section~\ref{reltab}, 
we determine with which fan it is best to start.

\begin{algorithm}[H]
\begin{algorithmic}[5]
\Require A list $F$ of fans $F_1,\dots F_N$ in $\RR^n$ where each $F_i$ is represented by a list of cones covering the support of $F_i$.
\Ensure A list of cones covering the support of $F_1\wedge\cdots\wedge F_N$.
\Procedure{DynamicEnumeration}{Cone $C$, Set $I$}
\If{$C\not=\emptyset$}
\If{$I=\emptyset$}
\State Output $C$
\Else
\State Greedily choose index $i\in I$.
\For{each cone $D$ in $F_i$}
\State {\sc DynamicEnumeration}($C\cap D$, $I\setminus\{i\}$)
\EndFor
\EndIf
\EndIf
\EndProcedure
\State {\sc DynamicEnumeration}($\RR^n$, $\{1,\dots,|F|\}$)
\end{algorithmic}
\caption{Dynamic enumeration}
\label{dynamic:recursive}
\end{algorithm}

Every intermediate cone intersection $C$
must be intersected with each remaining fan in order to contribute to the final common refinement, but it can be intersected with the
fans in any order. We greedily choose the next fan to intersect with $C$,
as a fan with which $C$ is expected to have few non-empty intersections. After intersecting with the fan we get a new set of cones, each contained within~$C$. 
Each of these new cones must be intersected with the remaining fans,
but this can happen in whatever order seems to be the most efficient,
determined through greedy selection. 
This process ends when $C$ is the result of an intersection of a cone from each fan.

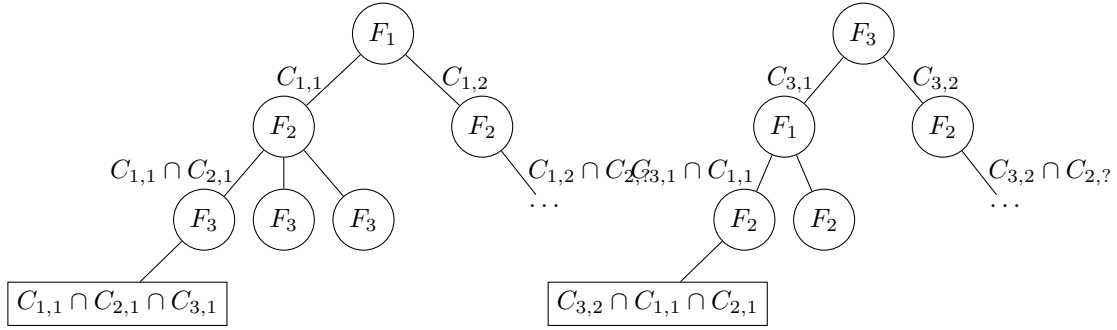
\begin{figure*}
\centering
\begin{minipage}[b]{.4\textwidth}
\centering
\begin{forest}
[$F_1$,circle, draw
     [$F_2$,circle, draw,edge label={node[midway,left] {$C_{1,1}$}}
       [$F_3$,circle, draw,edge label={node[midway,left] {$C_{1,1} \cap C_{2,1}$}} 
         [$C_{1,1} \cap C_{2,1} \cap C_{3,1}$,rectangle, draw]
         [,phantom]
         [,phantom]
       ]
       [$F_3$,circle, draw]
       [$F_3$,circle, draw]
     ]
     [$F_2$,circle, draw, edge label={node[midway,right] {$C_{1,2}$}}
     [,phantom]
     [,phantom]
     [,phantom]
     [$\ldots$, edge label={node[midway,right] {$C_{1,2} \cap C_{2,?}$}} ]]
     ]
 ]
\end{forest}

\end{minipage}\qquad
\begin{minipage}[b]{.4\textwidth}
\centering
\begin{forest}
[$F_3$,circle, draw
    [$F_1$,circle, draw,edge label={node[midway,left] {$C_{3,1}$}}
       [$F_2$,circle, draw,edge label={node[midway,left] {$C_{3,1} \cap C_{1,1}$}} 
         [$C_{3,2} \cap C_{1,1} \cap C_{2,1}$,rectangle, draw]
         [,phantom]
         [,phantom]
       ]
       [$F_2$,circle, draw]
     ]
     [$F_2$,circle, draw, edge label={node[midway,right] {$C_{3,2}$}}
     [,phantom]
     [,phantom]
     [,phantom]
     [$\ldots$, edge label={node[midway,right] {$C_{3,2} \cap C_{2,?}$}} ]]
     ]
]
\end{forest}
\end{minipage}\qquad
\caption{These trees illustrate the difference between static
and dynamic enumeration for three fans $F_1$, $F_2$, and $F_3$.
The left tree represents static enumeration, where the ordering
of the fans is established and does not change.
The right tree represents dynamic enumeration, where the starting
fan is greedily selected, and each cone greedily chooses which
fan to be intersected with next.}
\label{treediagram}
\end{figure*}

Algorithm~\ref{dynamic:recursive} and Algorithm~\ref{static:recursive}
have recursion trees with different shapes.
In the setting of mixed volume computation it was observed in \cite{MT08} that the tree of Algorithm~\ref{dynamic:recursive} has far fewer vertices. 
Figure~\ref{treediagram} demonstrates the difference
between the tree traversal of static enumeration and dynamic enumeration.

\subsection{Greedy Selection}\label{reltab}

The basic unit of work for Algorithm~\ref{static:recursive}
and Algorithm~\ref{dynamic:recursive}
is intersecting a pair of polyhedral cones. Every greedy choice
we make is done to minimize the number of necessary intersections
while avoiding adding an additional computational burden. To this end,
the choice of the first fan is easy: pick the fan with
the fewest cones.

The greedy metric used during the tree traversal is more difficult.
To facilitate the choices, we use relation tables
(introduced in~\cite{GL03}), tables that store whether or not pairs 
of cones could intersect. 
Before we define a relation table, it is useful to introduce 
additional notation: call $C_{i,j}$ the $j$th cone of fan
$T(P_i)$. For a cone $C$ of fan $T(P)$,
we define {\em the relation table $R(i,j)$} to be a boolean array
such that
\begin{equation}
R(i,j) = 
\begin{cases}
1, & \text{if } C \cap C_{i,j} \ne \emptyset \\
0, & \text{if } C \cap C_{i,j} = \emptyset \\
0, & \text{if } P = P_i \\
\end{cases}
\end{equation}
where $1 \le i \le \N \text{ and } 1 \le j \le \#\text{Edges}(P_i)$.
It is faster to check if $C \cap D = \emptyset$
than it is to compute $C \cap D$ with redundant inequalities removed. Checking if an
intersection is empty is equivalent to checking the feasibility
of a linear system.

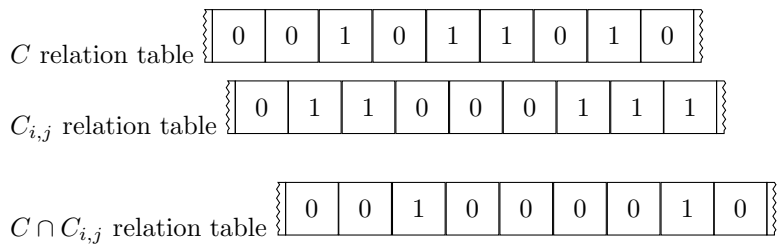
\begin{figure}[H]
$C$ relation table
\begin{tikzpicture}
  \tikzset{tape/.style={minimum size=.7cm, draw}}
  \begin{scope}[start chain=0 going right, node distance=0mm]
   \foreach \x [count=\i] in {0,0,1,0,1,1,0,1,0} {
    \ifnum\i=9 
      \node [on chain=0, tape, outer sep=0pt] (n\i) {\x};
      \draw (n\i.north east) -- ++(.1,0) decorate [decoration={zigzag, segment length=.12cm, amplitude=.02cm}] {-- ($(n\i.south east)+(+.1,0)$)} -- (n\i.south east) -- cycle;
     \else \ifnum\i=1 
      \node [on chain=0, tape, outer sep=0pt] (n\i) {\x};
      \draw (n\i.north west) -- ++(-.1,0) decorate [decoration={zigzag, segment length=.12cm, amplitude=.02cm}] {-- ($(n\i.south west)+(-.1,0)$)} -- (n\i.south west) -- cycle;
     \else
      \node [on chain=0, tape] (n\i) {\x};
     \fi \fi
   }
  \end{scope}
\end{tikzpicture}

\vspace{2mm}

$C_{i,j}$ relation table
\begin{tikzpicture}
  \tikzset{tape/.style={minimum size=.7cm, draw}}
  \begin{scope}[start chain=0 going right, node distance=0mm]
   \foreach \x [count=\i] in {0,1,1,0,0,0,1,1,1} {
    \ifnum\i=9 
      \node [on chain=0, tape, outer sep=0pt] (n\i) {\x};
      \draw (n\i.north east) -- ++(.1,0) decorate [decoration={zigzag, segment length=.12cm, amplitude=.02cm}] {-- ($(n\i.south east)+(+.1,0)$)} -- (n\i.south east) -- cycle;
     \else \ifnum\i=1 
      \node [on chain=0, tape, outer sep=0pt] (n\i) {\x};
      \draw (n\i.north west) -- ++(-.1,0) decorate [decoration={zigzag, segment length=.12cm, amplitude=.02cm}] {-- ($(n\i.south west)+(-.1,0)$)} -- (n\i.south west) -- cycle;
     \else
      \node [on chain=0, tape] (n\i) {\x};
     \fi \fi
   }
  \end{scope}
\end{tikzpicture}

\vspace{5mm}

$C \cap C_{i,j}$ relation table
\begin{tikzpicture}
  \tikzset{tape/.style={minimum size=.7cm, draw}}
  \begin{scope}[start chain=0 going right, node distance=0mm]
   \foreach \x [count=\i] in {0,0,1,0,0,0,0,1,0} {
    \ifnum\i=9 
      \node [on chain=0, tape, outer sep=0pt] (n\i) {\x};
      \draw (n\i.north east) -- ++(.1,0) decorate [decoration={zigzag, segment length=.12cm, amplitude=.02cm}] {-- ($(n\i.south east)+(+.1,0)$)} -- (n\i.south east) -- cycle;
     \else \ifnum\i=1 
      \node [on chain=0, tape, outer sep=0pt] (n\i) {\x};
      \draw (n\i.north west) -- ++(-.1,0) decorate [decoration={zigzag, segment length=.12cm, amplitude=.02cm}] {-- ($(n\i.south west)+(-.1,0)$)} -- (n\i.south west) -- cycle;
     \else
      \node [on chain=0, tape] (n\i) {\x};
     \fi \fi
   }
  \end{scope}
\end{tikzpicture}

\caption{Sample intersection of two relation tables. Each relation
table is an array of bits, so intersecting a pair of relation
tables is a bitwise AND.}
\label{relationtables}
\end{figure}

Before our algorithm begins,
we initialize each cone's relation table by intersecting each cone $C$
with every cone not in the same fan.
We store this information compactly on a bit array on each cone object.
When we intersect cone objects in Algorithm~\ref{dynamic:iterative},
not only do we intersect the polyhedral cones, but we also intersect
their associated relation tables. Intersecting relation tables
requires creating a new bit array that is equal to a bitwise AND of the two 
input bit arrays, as shown in Figure~\ref{relationtables}. As this is a
bit operation, it is very fast.

To make the greedy choices that our algorithm requires, our software
consults a cone's relation table to find the unused fan
with the fewest cones with which it could intersect. This greedy
selection requires minimal additional computation and leads to
large speedups as will be shown in Section~\ref{experimentalresults}.
An additional benefit of the relation tables is that they allow us to
avoid intersecting cones that are already known to not intersect. This is 
illustrated in Algorithm~\ref{dynamic:verbose}.

\begin{algorithm}[H]
\begin{algorithmic}[5]
\Require A list $F$ of fans $F_1,\dots F_N$ in $\RR^n$ where each $F_i$ is represented by a list of cones covering the support of $F_i$.
\Ensure A list of cones covering the support of $F_1\wedge\cdots\wedge F_N$.
\Procedure{DynamicEnumeration}{Cone $C$, Set $I$}
\If{$C\not=\emptyset$}
\If{$I=\emptyset$}
\State Output $C$
\Else
\State Choose index $i\in I$ such that $F_i$ has fewest
\State {\hspace{0.4cm}}cones which $C$ could intersect.
\For{each cone $D$ in $F_i$}
\If{$C$'s relation table allows $C\cap D\not=\emptyset$}
\State Intersect $C$'s relation table with $D$'s
\State {\hspace{0.4cm}}relation table, and store on $C\cap D$
\State {\sc DynamicEnumeration}($C\cap D$, $I\setminus\{i\}$)
\EndIf
\EndFor
\EndIf
\EndIf
\EndProcedure
\State Compute relation tables for $\RR^n$ and the cones in $F$
\State {\sc DynamicEnumeration}($\RR^n$, $\{1,\dots,|F|\}$)

\end{algorithmic}
\caption{Full dynamic enumeration algorithm}
\label{dynamic:verbose}
\end{algorithm}

\section{Parallel Implementation}

The problem of computing a tropical prevariety
can be seen as traversing a tree, 
which can run in parallel.

While the polytopes, fans, and cones in the tropical prevariety
all live in $\RR^n$, we point out that their defining data is exact,
spanned by points with {\em integer} coordinates. 
All our computations are performed with arbitrary precision
integer arithmetic.  
The rapid coefficient growth is not polynomial in~$n$, 
but $n$ is relatively small in our applications.

\subsection{Software Setup}

Our choices of software packages both enables and limits
our options for a parallel implementation.

We chose to use the Parma Polyhedra Library (PPL) for all
polyhedral computations~\cite{BHZ08}. We first call it to
find the vertices of the Newton polytopes,
and we later call it many times to compute intersections of
polyhedral cones.
PPL uses arbitrary precision integers during its computations via the GNU Multi Precision Arithmetic Library (GMP)~\cite{GMP}.
If we had chosen to use a library that
did not use arbitrary precision integers, the software
would need to exercise care and certify that there
were no incorrect answers due to floating point error.
Since computing a prevariety requires intersecting many
polyhedral cones in sequence, as the depth of the tree
increases, the likelihood of floating point error also increases.
Using a polyhedral library that uses GMP integers
avoids this challenge.

PPL has recently become threadsafe, which allows us to use it in
a parallel implementation. Unfortunately, if multiple threads
are intersecting pairs of polyhedra simultaneously, linear speedups
will not be achieved. This is due to the fact that when multiple
threads attempt to allocate GMP integers from the heap, a linear
speedup is not attained. Modest improvements in speedup come from using
the allocator TCMalloc~\cite{TCMalloc}.

\subsection{Design Considerations}

For computing a tropical prevariety,
we distinguish three stages in the algorithm.
For each of the three stages, we consider its parallel execution.

The algorithm begins by computing
the vertices of the Newton polytopes, which
is necessary to determine the normal fans.
Computing the vertices of distinct polytopes can be done in
parallel, but we have found it not to be necessary.
The polynomial systems of interest are sparse, with small
Newton polytopes spanned by relatively few monomials; if this
were not the case, determining the vertices would be
more difficult. In
our most computationally intensive benchmark, computing the vertices takes less than a second for a single thread.
Therefore, this component of the algorithm is not considered
to run in parallel.


The second stage we consider is the computation
of the relation tables.  Filling the relation tables
requires computing the intersection of many pairs of
polyhedral cones and testing if that intersection is non-empty.
This amounts to a job queue, where each job is the intersection
of two polyhedral cones.  The queue is filled with all of
the necessary jobs, then each process pops a job, records the
result of the intersection and returns to the queue. This
process continues until the job queue has been emptied.

The third stage of the algorithm 
leads to the greatest benefit:
developing a parallel version of the
recursion in Algorithm~\ref{dynamic:verbose}.
This will be addressed in the following sections.

\subsection{Coarse Grained Parallelism}

A first, coarse grained parallel version 
of Algorithm~\ref{dynamic:verbose} was implemented 
using forked processes, dividing the cones of the starting
fan among several processes.

Each process took its starting cones and performed
Algorithm~\ref{dynamic:verbose} on them, terminating when
finished. This approach was a natural starting point,
as it did not require communication between threads and it was
straightforward to implement. Since the processes were distinct, each
thread had its own heap, so we were closer to achieving linear
speedups in the polyhedral computations.
However, the time required for each process varied dramatically.
For a run of the cyclic-16 roots polynomial system with twenty threads,
the fastest threads finished in less than a day while the slowest thread
took more than three weeks to finish. This was an inefficient use
of resources, as a good parallel implementation uses all of a
computer's available resources for the duration of the computation.

\subsection{Work Stealing}

Our current parallel implementation applies work stealing~\cite{BL99},
using the run time parallel library provided by PPL.

The first barrier to creating a work stealing implementation
of the dynamic enumeration method is that
Algorithm~\ref{dynamic:verbose} is a recursive
algorithm, so it lacks a job queue.
We define a single job to be taking
a cone and intersecting it with the normal cones
of a polytope;
Algorithm~\ref{dynamic:iterative} transforms
Algorithm~\ref{dynamic:verbose} into an
algorithm with a work queue of these jobs.
This version of 
Algorithm~\ref{dynamic:verbose} will
find the prevariety with the same number
of cone intersections, but it will find the
cones of the prevariety in a different order.

\begin{algorithm}
\begin{algorithmic}[5]
\Require A list of fans $F_1,\dots F_N$ in $\RR^n$ where each $F_i$ is represented by a list of cones covering the support of $F_i$.
\Ensure A list of cones covering the support of $F_1\wedge\cdots\wedge F_N$.
\State Compute relation tables
\State $F$ := fan with fewest cones
\State Cones := Cones from $F$
\While{Cones$\not=\emptyset$}
\State $C$ := remove an element from Cones
\State Choose fan $F'$ not used to produce $C$ such that $F'$
\State {\hspace{0.4cm}} has fewest cones with which $C$ could intersect.
\For{each cone $D$ in $F'$}
\If{$C$'s relation table allows $C\cap D\not=\emptyset$}
\State Compute $C \cap D$
\If{$C \cap D \not=\emptyset$}
\If{$C \cap D$ used all fans}
\State Output $C \cap D$
\Else
\State Intersect $C$'s relation table with $D$'s
\State {\hspace{0.4cm}}relation table, and store on $C\cap D$
\State Add $C\cap D$ to Cones
\EndIf
\EndIf
\EndIf
\EndFor
\EndWhile
\end{algorithmic}
\caption{Iterative version of dynamic enumeration}
\label{dynamic:iterative}
\end{algorithm}

There are two benefits to finding
cones in the tropical prevariety quickly.
Once a cone in the prevariety has been
discovered, it can be printed to file,
so the memory that it consumed can
be freed. Additionally, when cones in
the prevariety are found, post-processing
can begin, which may vary depending on the
application. One application
of interest is computing power series
expansions of positive dimensional solution
sets of polynomial systems. Each cone
could lead to several distinct power
series, thus this process could
begin as soon as a single cone has
been found.

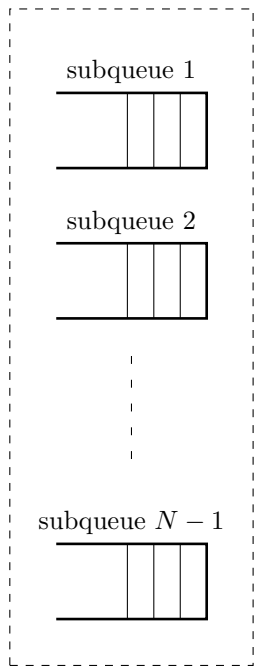
\begin{figure}
\begin{center}
\tikzset{
queuei/.pic={
  \draw[line width=1pt]
    (0,0) -- ++(2cm,0) -- ++(0,-1cm) -- ++(-2cm,0);
   \foreach \Val in {1,...,3}
     \draw ([xshift=-\Val*10pt]2cm,0) -- ++(0,-1cm);
   \node[above] at (1cm,0) {subqueue $#1$};   
  }
}
\begin{tikzpicture}[>=latex]
\path
  (2.5,3cm) pic {queuei=1}
  (2.5,1cm) pic {queuei=2}
  (2.5,-3cm) pic {queuei=\N-1};
\path
  (3.5,4cm) coordinate (aux1)
  (3.5,-4.5cm) coordinate (aux2)
  (2,0cm) coordinate (aux3)
  (5,0cm) coordinate (aux4);
\node[draw,dashed,text width=2.5cm,fit={(aux1) (aux2) (aux3) (aux4)}](dashed){};
\draw[loosely dashed] (3.5,-0.5) -- (3.5,-2);
\end{tikzpicture}
\caption{A queue made up of subqueues. Subqueue 1 contains 
cones from the starting polytope,
while subqueue $\N-1$ contains cones that
are that are the intersection of $\N-1$ cones.}
\label{jobqueue}
\end{center}
\end{figure}

To find cones in the prevariety as
quickly as possible, the job queue
is implemented as in Figure~\ref{jobqueue}, thereby also making it more stack-like.
When Algorithm~\ref{dynamic:iterative}
begins, cones are placed into
an initial subqueue, subqueue 1. When a cone from subqueue 1
is removed and has been intersected
with a set of cones from another polytope,
the resulting cones are put into subqueue 2.
To achieve the goal of finding cones in the
prevariety quickly, when choosing the next job, an
optimal strategy requires picking a cone from
the subqueue with the highest index.

To transform this algorithm into a
multi-threaded work stealing algorithm,
each thread must have its own queue
in the style of Figure~\ref{jobqueue}.
When picking a job to execute, a thread
first looks to its own queue and picks
a cone from the subqueue of highest index.
When a thread's queue is empty, it looks
to steal from another thread's queue,
but steals the job from the subqueue
of lowest index, as to require stealing
less often. Furthermore, if there are $j$
total threads, the $i$th thread
looks to steal from threads in the
following order:
$i+1, i+2, \ldots, j, 1, 2, \ldots i-1$.
This avoids having all threads attempting
to steal from the same thread, keeping
the theft spread out among different threads
and requiring fewer total robberies.

Since there is no communication between the various branches of the enumeration tree in Algorithm~\ref{dynamic:verbose}, an alternative to dealing with work stealing directly  is to phrase the recursive part of the algorithm as an abstract tree traversal and apply a general purpose parallel tree traverser. Thereby the parallelization aspects are entirely seperated from the problem domain. This approach was used in~\cite{Jen16} for the situation of mixed volume computation.

\section{Experimental Results}\label{experimentalresults}

For finding isolated solutions of polynomial systems,
there exist many standard benchmark problems. However,
few of these problems have positive dimensional
components as well, which makes them
inappropriate test cases for tropical prevarieties.
We will mention results from three standard benchmark problems with
positive dimensional solution components as well
as one problem from tropical geometry.

The code is available at \sloppy
\url{https://github.com/sommars/DynamicPrevariety}.
We compare it against Gfan which contains a single threaded and less efficient implementation of a variant of the dynamic enumeration algorithm. Except for the Gfan timings, all computations were done on a 2.2 GHz Intel Xeon E5-2699 processor
in a CentOS Linux workstation with 256 GB RAM using varying
numbers of threads.

\subsection{$n$-body and $n$-vortex Problems}

For equal masses, the central configurations in the classical $n$-body problem are solutions to the $\binom{n}{2}$ Albouy-Chenciner equations obtained by clearing denominators of the equations
\begin{equation}
\sum_{k=1}^n(x_{ik}^{-3}-1)(x_{jk}^2-x_{ik}^2-x_{ij}^2)
+(x_{jk}^{-3}-1)(x_{ik}^2-x_{jk}^2-x_{ij}^2)=0
\end{equation}
indexed by $i$ and $j$ where $1\leq i<j\leq n$, and there are $\binom{n}{2}$ pairwise distance variables $x_{12}\dots x_{(n-1)n}$ and $x_{ij}=x_{ji}$.

The $n$-vortex problem~\cite{HM09} arose from a generalization of a problem
from fluid dynamics that attempted to model vortex filaments. In this setting, the $\binom{n}{2}$ Albouy-Chenciner equations are obtained by clearing denominators of
\begin{equation}
\sum_{k=1}^n(x_{ik}^{-2}-1)(x_{jk}^2-x_{ik}^2-x_{ij}^2)
+(x_{jk}^{-2}-1)(x_{ik}^2-x_{jk}^2-x_{ij}^2)=0.
\end{equation}

\begin{table*}[h]
\begin{center}
  \begin{tabular}{r||r|r|r||r||r|r|r|}
 \multicolumn{1}{c||}{$n$}
& Static Enum & Dyn. Enum. & \#Rays & Gfan & 1 thread & 10 threads & 20 threads \\
 \hline
     4   &    114   &  114   & 2  & 0.020s  & 0.008s  & 0.017s  &  0.028s   \\ \hline
     5   &    682   &  676   &  0 & 0.058s  & 0.036s   & 0.053s &  0.073s   \\ \hline
     6   &    2,286   &  2,254   & 8  & 0.22s &  0.10s   &  0.11s &   0.16s   \\ \hline
     7   &    7,397   &  7,163   & 28 & 0.64s & 0.29s   &  0.26s &   0.37s \\ \hline
     8   &   19,619    &  18,315   &  94  & 2.87s &   0.79s   &  0.49s &   0.70s \\ \hline
     9   &   63,109    &  50,584   &  276 & 13.0s &  2.8s    &  1.2s &    1.4s  \\ \hline
     10  &  269,223     &  160,203   &  712 & 1m22s &  9.8s    &  4.4s &    3.7s  \\ \hline
     11  &   1,625,520    &  827,469   & 2,244    & 9m17s &   50s    &  16.8s  &    20.3s  \\ \hline
     12  &   11,040,912    &  5,044,441   & 5,582  & 82m33s &   5m2s    & 1m5s  &    1m3s   \\ \hline
     13  &       &  36,633,391   &  14,872  & &  46m59s &    8m30s  &     6m20s  \\ \hline
     14  &       &  264,463,730   &  49,114 &  &   6h22m56s   & 67m31s  &  46m37s     \\ \hline
     15  &       &  1,852,158,881   & 145,276  &     &   &  10h25m45s  & 7h43m57s \\ \hline
     16  &       &  13,715,434,028   &  527,126 &    &    &  84h20m37s  & 62h36m31s \\
  \end{tabular}
\caption{This table contains results from experiments with 
the cyclic-$n$ roots problem. The left half contains
the number of cone intersections
required in static enumeration and dynamic enumeration, as well
as the number of rays i.e. 1-dimensional cones in the produced fan.
The right half of the table contains timings of Gfan
and the dynamic prevariety software run with 1, 10, or 20 threads. The Gfan timings are for a single thread running on an Intel Xeon E2670 CPU. SoPlex~\cite{Wunderling1996} was enabled in the Gfan timings, providing a speed up of roughly a factor 3.}
\label{cyclic}
\end{center}
\end{table*}

Computing a tropical prevariety was essential in the argument of finiteness of the relative equilibria in the $4$-body problem~\cite{HM06}.
Tables~\ref{nbody} and~\ref{nvortex} contain data from
experiments with our implementation run on the $n$-body and $n$-vortex equations above.
Since the problems increase in difficulty quickly,
we can only compute few prevarieties in the family.

\begin{table}[h]
\begin{center}
  \begin{tabular}{c|r|r|r|}
    $n$ & \#Rays & 1 thread & 20 threads \\
 \hline
     3  & 4  & 0.014s & 0.038s  \\ \hline
     4  & 57  & 0.77s & 0.61s   \\ \hline
     5  & 2908   & 2m37s & 34s  \\
  \end{tabular}
\caption{$n$-body problem: number of rays
and timings of the new software run with 1 or 20 threads.
The 6-body problem did not terminate in two days when
run with 20 threads.}
\label{nbody}
\end{center}
\end{table}

\begin{table}[h]
\begin{center}
  \begin{tabular}{c|r|r|r|}
    $n$ &  \#Rays & 1 thread & 20 threads\\
 \hline
     3  &     4  & 0.011s & 0.03s  \\ \hline
     4  &     27  & 0.44s & 0.42s  \\ \hline
     5  &     643  & 30.1s & 10.9s \\ \hline
     6  &     152,514  &  & 3h16m13s \\
  \end{tabular}
\caption{$n$-vortex problem: number of rays
and timings of the new software run with 1 or 20 threads.
The 7-vortex problem did not terminate in two days when
run with 20 threads.}
\label{nvortex}
\end{center}
\end{table}

\subsection{$4 \times 4$ minors of a $5 \times 5$ matrix}

In~\cite{DSS03}, the authors pose the question of whether
or not the $4 \times 4$ minors of a $5 \times 5$ matrix
form a tropical basis. It was answered in the
affirmative~\cite{CJR11}, where one proof strategy
required computing the prevariety 
defined by the $4\times 4$ minors of the following matrix:

\begin{equation}
\left(\begin{array}{rrrrr}
x_{11} &  x_{12} & x_{13} & x_{14} & x_{15} \\
x_{21} &  x_{22} & x_{23} & x_{24} & x_{25} \\
x_{31} &  x_{32} & x_{33} & x_{34} & x_{35} \\
x_{41} &  x_{42} & x_{43} & x_{44} & x_{45} \\
x_{51} &  x_{52} & x_{53} & x_{54} & x_{55} \\
\end{array}\right).
\end{equation}

In~\cite{CJR11}, the $4\times 4$ minors prevariety
was computed in two weeks, using four threads, or
eight weeks of computation time if run
linearly. They also exploited symmetry of the problem,
which reduced the computation time by an expected factor between 2-10x.

The dynamic prevariety software ran in under ten hours
using twenty threads. However, these two trials cannot
fairly be compared, as processor speeds have increased
in the eight years since the computation was run
in~\cite{CJR11}.

\subsection{Cyclic-$n$ roots}

The cyclic $n$-roots problem asks for the solutions
of a polynomial system, commonly formulated as
\begin{equation} \label{eqcyclicsys}
   \begin{cases}
   x_{0}+x_{1}+ \cdots +x_{n-1}=0 \\
   i = 2, 3, \ldots, n-1: 
    \displaystyle\sum_{j=0}^{n-1} ~ \prod_{k=j}^{j+i-1}
    x_{k~{\rm mod}~n}=0 \\
   x_{0}x_{1}x_{2} \cdots x_{n-1} - 1 = 0. \\
\end{cases}
\end{equation}
This problem is important in the study of biunimodular vectors,
a notion that traces back to Gauss, as stated in~\cite{FR15}.
In~\cite{Bac89}, Backelin showed that if $n$ has a divisor that
is a square, i.e. if $d^2$ divides $n$ for $d \geq 2$, then
there are infinitely many cyclic $n$-roots.
The conjecture of Bj{\"{o}}rck and Saffari~\cite{BS95},
\cite[Conjecture~1.1]{FR15}
is that if $n$ is not divisible by a square, 
then the set of cyclic $n$-roots is finite.
If the dimension is a prime number, then the number of solutions
if finite, as proven in~\cite{Haa08} with an explicit count of the
number of solutions given.

The cyclic-$n$ roots problem scales slowly, so it is a good
case study to examine the effectiveness of
Algorithm~\ref{dynamic:iterative} in detail. The first
two columns of Table~\ref{cyclic} show that as the size of
the problem increases, the benefit of dynamic enumeration
increases as well.

From the right hand portion of Table~\ref{cyclic}, it 
can be seen that there is a speedup as the number of threads
increases. In cyclic-14, a speedup of 5.67 was achieved with
ten threads while a speedup of 8.21 was achieved with twenty
threads. These speedups are not linear, due to
the GMP integer allocation issue. However, computing in parallel
dramatically reduces computation time, so it is beneficial
in practice.

As the number of cones increase, postprocessing the results
becomes increasingly difficult. For cyclic-16, there are
many cones of high dimension, which makes the prevariety
more challenging to compute, see Table~\ref{cyclic16maximalcount}.
\begin{table}[h]
\begin{center}
  \begin{tabular}{c|r}
    Dim. &  \#Maximal Cones\\
 \hline
     1  &   0    \\ \hline
     2  &   768    \\ \hline
     3  &   114,432    \\ \hline
     4  &   1,169,792    \\ \hline
     5  &   1,007,616    \\ \hline
     6  &   2,443,136    \\ \hline
     7  &   4,743,904    \\ \hline
     8  &   109,920    \\ 
  \end{tabular}
\caption{Number of maximal cones in the prevariety of cyclic-16
by dimension.}
\label{cyclic16maximalcount}
\end{center}
\end{table}


\end{document}